\documentclass[runningheads,a4paper]{llncs}
\usepackage{graphicx} 
\usepackage{url}

\begin{document}

\title{Spatio-temporal Models \\
 for Formal Analysis and Property-based Testing}

\author{Nasser Alzahrani, Maria Spichkova, Jan Olaf Blech}
\institute{RMIT University, Melbourne, Australia \\  
\email{s3297335@student.rmit.edu.au, \{maria.spichkova,janolaf.blech\}@rmit.edu.au} }

\newcommand{\epar}[1]{``#1''}
\newcommand{\focust}{\textsc{Focus}$^{ST}$}

\maketitle

\begin{abstract} 
This paper presents our ongoing work  on spatio-temporal models for formal analysis and property-based testing. Our proposed framework aims at reducing the impedance mismatch between formal methods and practitioners. We introduce a set of formal methods and explain their interplay and benefits in terms of usability.\footnote{%
Preprint. Accepted to the Software Technologies: Applications and Foundations (STAF 2016). Final version published by Springer International Publishing AG.
}

\end{abstract}
%========================================
\section{Introduction}

Specifying safety-critical systems,   
it is not enough to use controlled languages and semiformal languages -- 
the precise and easy-to-read formal specification  is essential to ensure that the safety properties of the system really hold. 
Moreover, the software development process should include aspects of human factors engineering, to 
 improve the quality of software and to deal with human factors in a systematic way, cf.  \cite{ICSE_2015_HF}.
Human factor aspects usually cover the design of human-computer interface of the software, 
human-related aspects of the development process, as well as the corresponding automatisation. 
By the Engineering Error Paradigm \cite{RedmillRajan},
humans are seen as a \epar{component of the system} (almost equivalent to software and hardware components in the sense of operation with data and other components), which is the most unreliable in the system. 

Software errors can cause wasting of resources \cite{patra2007worst,charette2005software}. 
An estimate of one trillion US dollars was spent on IT hardware, software and services by governments around the world.
Software errors can also be fatal, and in many cases they might be prevented by having a more human-oriented development process and methods. 
As per statistics presented by Dhillon \cite{Dhillon}, humans are responsible for 30\% to 60\% the total errors which directly or indirectly lead to the accidents, and in the case of aviation and traffic accidents, 80\% to 90\% of the errors were due to humans. 
Thus, it is necessary to have human factors engineering as a part of the software development process.
One of the widely cited accidents in safety-critical systems are the accidents involved massive radiation overdoses by the Therac-25 
(a radiation therapy machine used in curing cancer) 
that lead to deaths and serious injuries of patients which received thousand times the normal dose of radiation \cite{Miller1987,leveson1993investigation}. 
The causes of these accidents were software failures as well as problems with the system interface. 
The error was improbable to reproduce because it required very specific sequence of
commands in order to occur. 
The improbability of the sequence makes the error unlikely to be noticed with manual testing because it is almost impossible to think of all combinations of commands and edge cases. Automatisation might solve this problem, but the challenge is to create 
an automatisation which is not only efficient but also easy-to-use, i.e., is human-oriented.

One of the challenges in software engineering is to develop correct software. The software should meet user requirements, its properties should satisfy the model corresponding to
design objective and the implementation should pass all functional tests. 
Rigorous reasoning is the only way to avoid subtle errors in algorithms, and it should be as simple as possible by making the 
underlying formalism simple tools \cite{lamport_hybrid_1993}.
Formal methods (FMs) refer to a class of mathematical techniques used in development of large scale complex systems. These 
techniques can result in high-quality systems that can be implemented on-time, within budgets and satisfy user requirements \cite{bowen1995seven}.

The value of FMs in real systems has far reaching consequences. For instance, FMs help engineers get the code right by getting the design right in the first place. Secondly, FMs help engineers gain a better understanding of the design.  
Despite all advantages, formal methods are not widely used in large-scale industrial software projects for many reasons \cite{enase2016cfm}. 
One of the core obstacles is the lack of readability and usability.
The syntax of FMs is often too complicated and unreadable for novices, which makes an impression that all the FMs 
require huge amount of training. 
There also is a prejudice that the return of investment is very minimal and only justified in critical systems such as medical 
devices, what is generally not true \cite{Newcombe2015}.

Spatio-temporal aspects of safety-critical systems are crucial to verify and to test a system, 
as in most cases the system properties should be analysed in relation to the time and to the location. 
To analyse spatio-temporal phenomena, we have to specify the corresponding spatial, temporal and event semantics formally and in a human-oriented way. 
The goal of our work is to increase usability of the analysis (in the sense of verification and testing) 
of the  spatio-temporal aspects on the base of the corresponding formal models. 
 
\emph{Property based testing}  allows us to generate huge numbers of system operations (e.g API calls or external events) and permute these operations in ways that is difficult for humans to think of. 
These combinations are then used to verify the system under test according to the spatio-temporal  specification.

\emph{Contributions:} 
The proposed framework
will help to reduce the impedance mismatch between formal methods and model-based representations and system code,  which in turn will help in increasing the adoption rate by practitioners. 
Our framework aims at providing a set of application programming interfaces (APIs) to map  programming language constructs to the formal methods representation.
The usability of formal methods will be improved indirectly, as the formal method constructs will be expressed in terms of system code.

%========================================
\section{Background}
\label{sec:related} 
\subsection{Formal Methods}

Formal methods were introduced as a means of clearly specifying system requirements. Hinchey \cite{hinchey_confessions_2003} argues 
that although formal methods are essential in the development of critical systems, 
they have not achieved the level of acceptance, nor level of use, that many believe they should. The uptake of formal methods 
has been far from ideal because many still believe that formal methods are difficult to use and require great mathematical 
expertise \cite{hinchey_confessions_2003}.  Spichkova reports \cite{hffm_spichkova}
  that in many cases simple changes of a specification method can make it more 
understandable and usable. She argues that such a simple kind of optimisation is often overlooked just because of its 
obviousness, and it would be wrong to ignore the possibility to optimise the language without much effort. 
For example, simply adding an enumeration to the formulas in a large formal specification makes its validation on the level of specification and 
discussion with cooperating experts much easier. 

Hinchey \cite{hinchey_confessions_2003} also assert that in addition to the benefits of abstraction, clarification, and disambiguation, using formal methods at the formal specification
level are invaluable documentation that greatly assist future system maintenance. This research incorporates specifications used in 
property-based testing to further help in precisely documenting the system.

Lamport \cite{lamport_hybrid_1993} states two reasons for using formal methods formulas instead of programming language tailored to the
specific problem:
\begin{itemize}
  \item Specialized languages often have limited realms of applicability. A language that permits a simple specification 
    for one system require a very complicated one for a different kind of system.
    The Duration Calculus seems to work well for real-time properties; but it cannot express simple 
    liveness properties. A formalism like TLA+ that, with no built-in primitives for real-time systems or procedures, 
    can easily specify gas burner for example, it is not likely to have difficulty with a different kind of gas burner.  
  \item Formalisms are easy to invent. However,  practical methods must have a precise language and robust tools.

\end{itemize}

\noindent
There are many examples where applying formal methods has lead to increasing reliability of systems. 
For example, a model checker  TLC 
was developed for TLA formula  was used to find errors in the cache coherence protocol for a Compaq multiprocessor \cite{yu1999model}. 
In addition, \cite{bowen1995seven} includes many examples of successfully using formal methods to design systems.

%------
\subsection{Property-Based Testing}
There are many styles in testing software. 
One popular style is that of \emph{example based testing}. 
In this style, test cases requires one to provide an example scenario for each feature. 
That is, each example may exercise one feature of the system under test and 
the test runs only once with relevant input. 
Dually, \emph{property based testing} allows for the use of randomly generated tests based
on system properties to test systems against their specifications and one test can run hundreds of times with different input values.
An example of such library in Haskell programming language is \emph{QuickCheck}. 
Hughes (inventor of \emph{QuickCheck}) showed that using this library allowed him to discover 
hundreds of bugs in critical systems such as automobiles and the DropBox file sharing service \cite{claessen_quickcheck:_2011}. 
However, \emph{QuickCheck} uses Haskell programming language specific constructs (such as arrays, integers) and more complicated 
data types (such as algebraic data types) to model the specification of a system. Therefore, this research will 
investigate the possibility to have formal models (BeSpaceD, TLA+ or \focust\  formulas) as specifications instead of Haskell constructs, 
as well as applicability of this approach for property based testing of real systems. 

Hughes \cite{hu2015functional} asserts that Dijkstra was wrong when he claimed that testing can never demonstrate the absence of bugs in software,
only their presence. 
Hughes argues that if we test properties that completely specify a function (such as the properties of reversing a list)
then property based testing will eventually find every possible bug. 
In practice this is not true, since we usually do not have a 
complete specification, but this style of testing is very effective in exploring scenarios that no human can think of
trying.

\emph{QuickCheck} started as a testing framework for testing pure functional programs \cite{claessen_quickcheck:_2011}. However, recent development in the area of property-based testing \cite{gerdes_linking_2015,hughes2010software} incorporates the state-fulness of systems. That allowed for the testing of state-ful systems and even test programs written in 
imperative languages such as C. 
Hughes assert that testing state-ful systems is challenging. He argues that the state is an implicit 
argument to and result from every API call, yet it is not directly accessible to the test code. Therefore, his solution was to 
model the state abstractly and introduce state transition function that model the operations in API under test.

However, the state transition in \emph{QuickCheck} is modelled manually using \emph{pre}, \emph{post} and \emph{next} functions for every operation in the system under test. On the other hand, our framework will generate these transitions automatically using  specification formulas.

%%============================================
\section{Proposed Framework}
\label{sec:framework}

\begin{figure}[!h]
  \centering
   \includegraphics[scale=0.38]{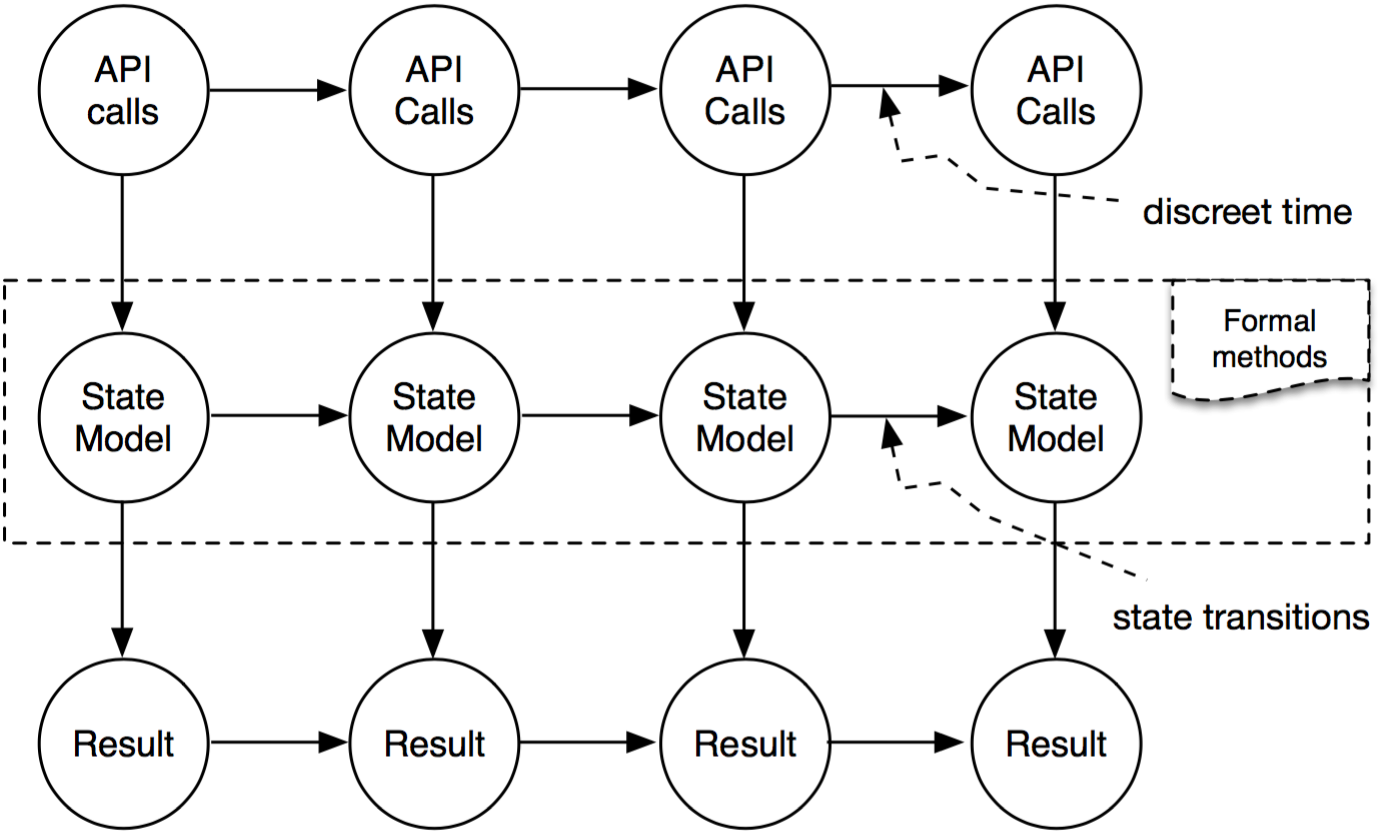}
  \caption{Proposed Framework}
  \label{fig:proposed_model}
\end{figure}

Figure \ref{fig:proposed_model} depicts the proposed model that will allow for combining formal methods with property-based-testing. The 
first row (API calls) represents the actual system under test. The second row represents the world in which the specification formulas lives. The time between subsequent API calls is modelled through a function of discreet time. Time functions are mapped to the corresponding state transitions between states. The general idea is to start with specifying the system using human-oriented modelling techniques founded on formal methods. Then,
to develop system software according to the specifications. Finally, to run the test suite to verify that the system
runs according to the specification. If a test fails, it will be the judgment of the engineer to decide whether
the errors were in the system software or in the specification formulas for which the system was not correctly specified. 

The implementation language of choice is Scala programming language. It was selected for many reasons. First of all, it is one of the most 
popular languages on the Java virtual machine. The ecosystem will make it possible to find quick answers for questions that are related
to technical aspects. Secondly, BeSpaceD is implemented in Scala. This will lower the impedance mismatch between research model and 
BeSpaceD. Finally, Scala, is a functional language. This will make working with the concepts of property based testing more natural 
and simple.  

For the property-based testing, we are going to apply the \emph{ScalaCheck library}.
However, since the research will investigate the substitution of the simplistic state machine in ScalaCheck with formal methods, the use of this library might be limited.

To relate the different modeling and abstraction layers to each other in the proposed framework, we are using 
 category theory. % 
Category theory helps in illuminating the relations of many aspects of the proposed ingredients that would be
unseen otherwise. Figure~\ref{fig:proposed_model} relates the human actions (API call), system states (state model) and results to each other. Our formal methods-based techniques will only be applied to the State-model level. This will help to stair the direction of future investigation of the proposed model.

%%============================================

\section{Initial set of Modeling Languages and Tools}

To create the initial set of  formal methods-based modeling languages and tools, we have selected the following ingredients, which have a number of similarities in syntax and
semantics and are also covering spatio-temporal aspects of the specifications:
\begin{itemize}
  \item 
  TLA+: Temporal logic of actions (TLA) is a logic developed by Leslie Lamport, 
    which combines temporal logic with a logic of actions. It is used to describe behaviours of concurrent systems, cf. \cite{lamport_temporal_1994}.
  \item 
  \focust: Formal language providing  concise but easily understandable specifications that is focused on  timing and spatial aspects of the system behaviour, cf. \cite{spichkova2014modeling,spichkova2007specification}.
  \item 
  BeSpaceD: A framework for modelling and checking behaviour of spatially distributed component systems, cf. \cite{blech_example_2015,blech_bespaced:_2014}.
\end{itemize}

The \focust\ language was inspired by Focus \cite{focus}, a framework for formal specification and development of interactive systems. 
In both languages, specifications are based on the notion of streams. 
However, in the original Focus input and output 
streams  of  a  component  are  mappings  of  natural  numbers to single messages,whereas a \focust stream is a mapping from natural numbers
to lists of messages within the corresponding time intervals. Moreover, the syntax of \focust is particularly devoted to
specify spatial (S) and timing (T) aspects in a comprehensible fashion, which is the reason to extend the name of the language by ST.
The \focust\ specification layout also differs from the original one: it is based on human factor analysis within formal methods \cite{hffm_spichkova,Spichkova2013HFFM}.

Design goals of BeSpaceD   include:
\begin{itemize}
  \item Ability to model spatial behaviour in a component oriented, simple and intuitive way
  \item Automatically analyse and verify systems and integration possibilities with other modelling and verification tools. 
\end{itemize}
Blech and Schmidt proposed a process for checking properties of 
models and described the approach using different examples \cite{blech_bespaced:_2014}. 
In our current work, we only focus on the spatio-temporal aspects of BeSpaceD.

From a programming language perspective, we create BeSpaceD models by using Scala case classes. During the specification process, this gives a functional abstract datatype feeling with a domain specific language flavour.
A typical BeSpaceD formula is shown below
{\small
\begin{verbatim}
 IMPLIES(AND(TimeInterval(300,605),Owner("AreaOfInterest")),
     OccupyBox(1051,3056,1505,3603))
\end{verbatim}
}
The language constructs comprise basic logical operators (such as {\tt AND} and {\tt IMPLIES}). Furthermore special constructs for space, time, and topology are incorporated.
In the example, {\tt OccupyBox} represents a rectangular two-dimensional space  while constructs such as {\tt Time\-Interval} allow for the modeling of temporal aspects possible.
A variety of different operators exist which facilitates the reasoning about geometric and topological constraints. Furthermore, connections to data sources from cyber-physical systems exists (e.g., lego-trains \cite{HOBH:16} and event analysis for industrial automation facilities \cite{7301533}) which facilitates the construction of demonstrators and conduction of experiments.

In our work we are using \focust\  and TLA+ for modelling the behaviour of systems, whereas the BeSpaceD functionality is invoked at a lower level to check and test properties of the specified systems.

To understand the workflow of the proposed model, we use the example of Therac25 mentioned in the introduction.
The machine included VT-100 terminal which controlled the PDP-11 computer.
The sequence of user actions leading to the accidents was as follows: 
\begin{itemize}
  \item user selects 25 MeV photon mode
  \item user enters ``cursor up''
  \item user select 25 MeV Electron mode
  \item previous commands have to take place in eight seconds
\end{itemize}

Therefore, we use algebraic data types to model the operations of the machine. Then we provide formal specification formulas and feed them to the framework. 

{\small
\begin{verbatim}
sealed abstract class Operation
case object CursorUp extends Operation
case object Select25MevPhotonMode extends Operation
case object Select25MevElectronMode extends Operation
case object OtherKindOfOperation extends Operation

type Therac25 = Sut

val init: TLAInit =  {.. some predicate ...}
val next: TLANext = {.. another predicate ...}

val correctBehaviours: List[TLAState] = 
   Therac25.correctBehaviours(init, next) 
Therac25.checkAgainst(correctBehaviours, randoms(Operation))

\end{verbatim}
}
 
The framework would generate large number of \emph{Operation} combinations that are more likely to catch the error that caused the fatal accidents. Frequencies of generated commands can be tailored to match real system behaviour. The example used TLA+ formulas. However, \focust\  formulas could have been used instead to specify the system. 

To achieve that, we have partially implemented the code that is responsible to generate  random  BespaceD constructs using techniques from functional programming. The Invariant generator is composed of smaller generators such as integer and string generators as shown in the code  below:

{\small
\begin{verbatim}
  trait Generator[+T] {
    self => 
  
    def generate: T
    
    def map[U](f: T => S): Generator[U] = new Generator[U] {
    	def generate = f(self.generate)
    }
  }
  
  val integers = new Generator[Int] {
     def generate = scala.util.Random.nextInt()
  }
  
  val booleans = integers.map(_ >= 0)
  
  val strings = integers.map(_.toString)
  
  def bSpaceD: Generator[Invariant] = for {
      int1  <- integers
      int2  <- integers
      int3  <- integers
      int4  <- integers
      int5  <- integers
      str   <- strings
  } yield IMPLIES(AND(TimeInterval(int1, int2),Owner(str)),
   			  OccupyBox(int3, int4, int5, int6))
 
\end{verbatim}
}

%%============================================
\section{Evaluation}

The evaluation is based on a case study that involves robotics that are installed in the Virtual Experiences Lab(VXLab) at RMIT University, Australia. 

\begin{figure}[!h]
  \centering
  \includegraphics[width=0.8\textwidth]{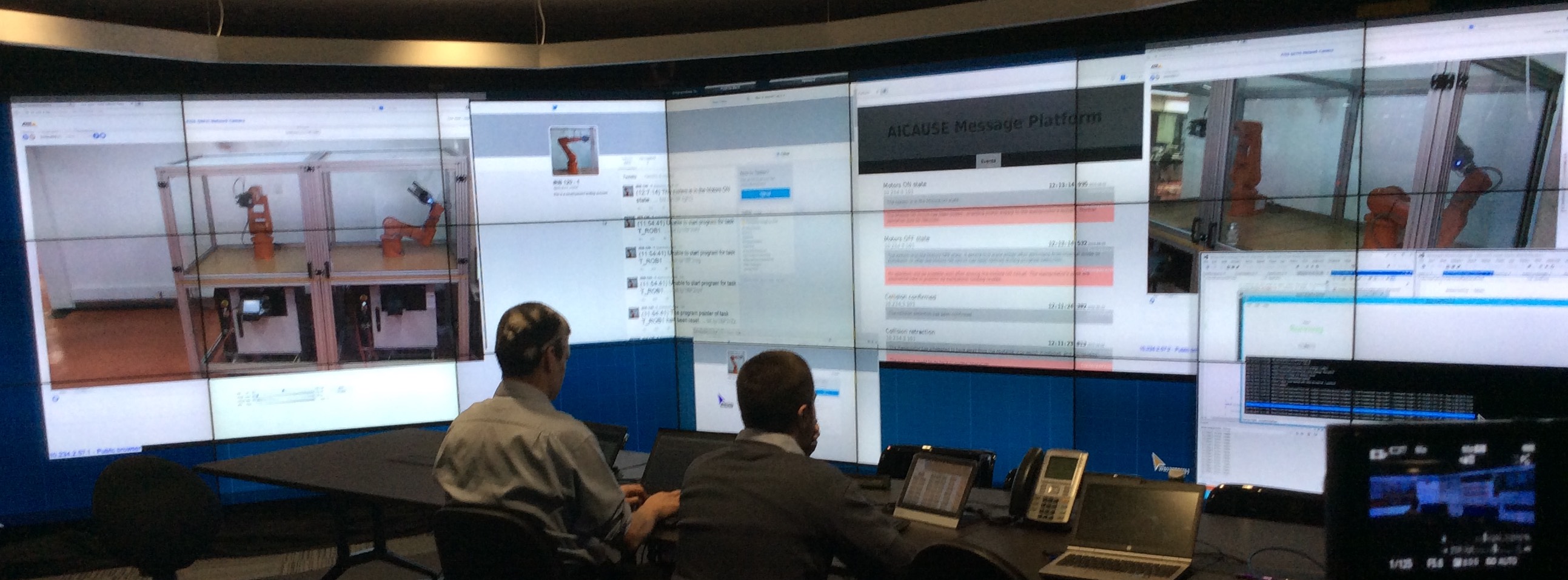}
  \caption{Interacting with robots from the VXLab at RMIT}
  \label{fig:vx:lab}
\end{figure}

The implemented model will be installed in the robotic arms or simulations of them. For instance, assuming the existence of the function \emph{initialisePosition(): Future[Position]} which is responsible to move a robotic arm to an initial position. The \emph{Future} data type is used because moving arms takes long time and we need to verify  the final position the arm reached after the API call. However, since \emph{initialisePosition()} is just returning the initial position, it will return instantly. The framework will call this API function and simultaneously check whether it is in accordance to the specified state. Failing tests for the intended
framework might indicate:
\begin{itemize}
  \item Failure in the software of the system under test. This is one of the benefits of property based testing. The found error may
    have never been discovered otherwise.
  \item Wrong specification. The system under test may have been wrongly underspecified. In this case, the engineer might change 
    the formulas to reflect system required properties.
\end{itemize}

Therefore, the input to the framework is formal-methods formulas and the output is the correct behaviours specified by these formulas. The formulas are written in host programming language (Scala in this research). For example, the initial state for the aforementioned robotic example would be specified as follows:
 {\small
\begin{verbatim}
val position: TLAVariable = TLAVariable("Y")
val init: TLAInit =  position
\end{verbatim}
}
For this simple example (the \emph{next} formula has been omitted for simplicity), the only possible correct behaviour for this specification formula is that \emph{position} should equal to "Y". The framework will then check whether the position was indeed "Y" after the call to \emph{initialisePosition()}, otherwise, it reports an error. 

\begin{table}[htp]
\caption{Evaluating cases with TLA+ Init Formulas}
\begin{center}
\begin{tabular}{|c c c c|}
\hline
 API Code & Init Formula & Result & Error?\\
\hline
\hline
\emph{initialisePosition()}	  & TLAVariable("Y") & "Y" & No
\\
\hline
\emph{initialisePosition()}	  & TLAVariable("Y") & "K" & Yes
\\
\hline
\emph{moveToQ()}	  &  TLAVariable("Q") & "Q" & Yes
\\
\hline
\emph{moveToR()}	  &  TLAVariable("Q")& "M" & Yes
\\
\hline

\end{tabular}
\end{center}
\label{tab:table2}
\end{table}

Table \ref{tab:table2} shows some examples for the evaluation of the intended framework using TLA+ formula (\focust\  evaluation will follow similar pattern). The first call to \emph{initialisePosition()} is correctly specified and the actual result  reflects the specification (assuming arm initial position is "Y"), as a result, it is regarded as a successful case. The second call to \emph{initialisePosition()} is different from the actual position, therefore, its was reported as an error. Although the result is expected for the call to  \emph{moveToQ()} in the third case, the framework reports an error because the specification is not correct (the arm can not logically move to its current position). Finally, \emph{moveToR} is reported as error because the actual result (reached position) is not correct. The result column is calculated by getting the value from the \emph{Future} dataype that each API call returns through \emph{onComplete} callback as follows:
{\small
\begin{verbatim}

initialisePosition() onComplete {
  case Success(position) => println(position)
  case Failure(t) => println("An error has occured: " + t.getMessage)
}
 
\end{verbatim}
}

%============================================
\section{Conclusions}
\label{sec:conclusions}

In this paper, we have presented ongoing work on the use of spatio-temporal models for formal methods-based analysis and testing. We have described different ingredients and their interplay: testing frameworks, TLA+, \focust\  and BeSpaceD. The overall goal of our research is the reduction of the impedance mismatch between formal methods and practitioners.

\bibliographystyle{abbrv}

\begin{thebibliography}{10}

\bibitem{7301533}
J.~Blech, I.~Peake, H.~Schmidt, M.~Kande, A.~Rahman, S.~Ramaswamy, S.~Sudarsan,
  and V.~Narayanan.
\newblock {Efficient Incident Handling in Industrial Automation through
  Collaborative Engineering}.
\newblock In {\em IEEE 20th Conference on Emerging Technologies Factory
  Automation (ETFA)}. IEEE Computer, Sept 2015.

\bibitem{blech_example_2015}
J.~O. Blech.
\newblock An example for {BeSpaceD} and its use for decision support in
  industrial automation.

\bibitem{blech_bespaced:_2014}
J.~O. Blech and H.~Schmidt.
\newblock {BeSpaceD}: Towards a tool framework and methodology for the
  specification and verification of spatial behavior of distributed software
  component systems.

\bibitem{bowen1995seven}
J.~P. Bowen and M.~G. Hinchey.
\newblock Seven more myths of formal methods.
\newblock {\em IEEE software}, 12(4):34, 1995.

\bibitem{focus}
M.~Broy and K.~St{\o}len.
\newblock {\em Specification and Development of Interactive Systems: Focus on
  Streams, Interfaces, and Refinement}.
\newblock Springer, 2001.

\bibitem{charette2005software}
R.~N. Charette.
\newblock Why software fails [software failure].
\newblock {\em Spectrum, IEEE}, 42(9):42--49, 2005.

\bibitem{claessen_quickcheck:_2011}
K.~Claessen and J.~Hughes.
\newblock {QuickCheck}: A lightweight tool for random testing of haskell
  programs.
\newblock 46(4):53--64.

\bibitem{Dhillon}
B.~Dhillon.
\newblock {\em Engineering Usability: Fundamentals, Applications, Human
  Factors, and Human Error}.
\newblock American Scientific Publishers, 2004.

\bibitem{gerdes_linking_2015}
A.~Gerdes, J.~Hughes, N.~Smallbone, and M.~Wang.
\newblock Linking unit tests and properties.
\newblock In {\em Proceedings of the 14th {ACM} {SIGPLAN} Workshop on Erlang},
  Erlang 2015, pages 19--26. {ACM}.

\bibitem{hinchey_confessions_2003}
M.~G. Hinchey.
\newblock Confessions of a formal methodist.
\newblock In {\em Proceedings of the Seventh Australian Workshop Conference on
  Safety Critical Systems and Software 2002 - Volume 15}, {SCS} '02, pages
  17--20. Australian Computer Society, Inc.

\bibitem{HOBH:16}
S.~Hordvik, K.~{\O}seth, J.~O. Blech, and P.~Herrmann.
\newblock {A Methodology for Model-based Development and Safety Analysis of
  Transport Systems}.
\newblock In {\em 11th International Conference on Evaluation of Novel
  Approaches to Software Engineering (ENASE)}, 2016.

\bibitem{hu2015functional}
Z.~Hu, J.~Hughes, and M.~Wang.
\newblock How functional programming mattered.
\newblock {\em National Science Review}, 2(3):349--370, 2015.

\bibitem{hughes2010software}
J.~Hughes.
\newblock Software testing with quickcheck.
\newblock In {\em Central European Functional Programming School}, pages
  183--223. Springer, 2010.

\bibitem{lamport_hybrid_1993}
L.~Lamport.
\newblock Hybrid systems in {TLA}+.
\newblock In R.~L. Grossman, A.~Nerode, A.~P. Ravn, and H.~Rischel, editors,
  {\em Hybrid Systems}, number 736 in Lecture Notes in Computer Science, pages
  77--102. Springer Berlin Heidelberg.
\newblock {DOI}: 10.1007/3-540-57318-6\_25.

\bibitem{lamport_temporal_1994}
L.~Lamport.
\newblock The temporal logic of actions.
\newblock 16(3):872--923.

\bibitem{leveson1993investigation}
N.~G. Leveson and C.~S. Turner.
\newblock An investigation of the therac-25 accidents.
\newblock {\em Computer}, 26(7):18--41, 1993.

\bibitem{Miller1987}
E.~Miller.
\newblock {The Therac-25 Experience}.
\newblock In {\em Conf. State Radiation Control Program Directors}, 1987.

\bibitem{Newcombe2015}
C.~Newcombe, T.~Rath, F.~Zhang, B.~Munteanu, M.~Brooker, and M.~Deardeuff.
\newblock How amazon web services uses formal methods.
\newblock {\em Communications ACM}, 58(4):66--73, Mar. 2015.

\bibitem{patra2007worst}
S.~Patra.
\newblock Worst-case software safety level for braking distance algorithm of a
  train.
\newblock In {\em System Safety, 2007 2nd Institution of Engineering and
  Technology International Conference on}, pages 206--210. IET, 2007.

\bibitem{RedmillRajan}
F.~Redmill and J.~Rajan.
\newblock {\em Human factors in safety-critical systems}.
\newblock {Butterworth-Heinemann}, 1997.

\bibitem{hffm_spichkova}
M.~Spichkova.
\newblock {Human Factors of Formal Methods}.
\newblock In {\em {In IADIS Interfaces and Human Computer Interaction 2012}}.
  IHCI 2012, 2012.

\bibitem{Spichkova2013HFFM}
M.~Spichkova.
\newblock {\em Design of formal languages and interfaces: ``Formal'' does not
  mean ``unreadable''}.
\newblock IGI Global, 2013.


\bibitem{spichkova2014modeling}
M.~Spichkova, J.~O. Blech, P.~Herrmann, and H.~W. Schmidt.
\newblock Modeling spatial aspects of safety-critical systems with Focus$^{ST}$.
\newblock In {\em MoDeVVa@ MoDELS}, pages 49--58. Citeseer, 2014.

\bibitem{spichkova2007specification}
M.~Spichkova.
\newblock {\em Specification and seamless verification of embedded real-time
  systems: FOCUS on Isabelle}.
\newblock PhD thesis, Technical University Munich, 2007.

\bibitem{ICSE_2015_HF}
M.~Spichkova, H.~Liu, M.~Laali, and H.~W. Schmidt.
\newblock Human factors in software reliability engineering.
\newblock {\em Workshop on Applications of Human Error Research to Improve
  Software Engineering (WAHESE2015)}, 2015.

\bibitem{yu1999model}
Y.~Yu, P.~Manolios, and L.~Lamport.
\newblock Model checking TLA+ specifications.
\newblock In {\em Correct Hardware Design and Verification Methods}, pages
  54--66. Springer, 1999.

\bibitem{enase2016cfm}
A.~Zamansky, G.~Rodriguez-Navas, M.~Adams, and M.~Spichkova.
\newblock Formal methods in collaborative projects.
\newblock In {\em {11th International Conference on Evaluation of Novel
  Approaches to Software Engineering}}. IEEE, 2016.

\end{thebibliography}

\end{document}